\documentclass[amsmath,amssymb,nofootinbib,notitlepage,superscriptaddress,twocolumn]{revtex4-1}
\pdfoutput=1

\usepackage{aas_macros,mathtools}
\usepackage[bottom]{footmisc}
\usepackage[colorlinks=true]{hyperref}
\usepackage[raggedright]{subfigure}
\let\originalleft\left
\let\originalright\right
\renewcommand{\left}{\mathopen{}\mathclose\bgroup\originalleft}
\renewcommand{\right}{\aftergroup\egroup\originalright}

\newcommand{\ab}[1]{\left|#1\right|}
\newcommand{\br}[1]{\left[#1\right]}
\newcommand{\cu}[1]{\left\{#1\right\}}
\newcommand{\pa}[1]{\left(#1\right)}
\newcommand{\ed}{\mathop{}\!\mathrm{d}}
\newcommand{\pd}{\mathop{}\!\partial}
\def\lsim{\mathrel{\raise.3ex\hbox{$<$\kern-.75em\lower1ex\hbox{$\sim$}}}}
\def\gsim{\mathrel{\raise.3ex\hbox{$>$\kern-.75em\lower1ex\hbox{$\sim$}}}}

\DeclareMathOperator\sign{sign}

\begin{document}

\title{Universal Polarimetric Signatures of the Black Hole Photon Ring}

\author{Elizabeth Himwich}
\email{himwich@g.harvard.edu}
\affiliation{Center for the Fundamental Laws of Nature, Harvard University, Cambridge, MA 02138, USA}
\affiliation{Black Hole Initiative at Harvard University, 20 Garden Street, Cambridge, MA 02138, USA}
\author{Michael D. Johnson}
\email{mjohnson@cfa.harvard.edu}
\affiliation{Black Hole Initiative at Harvard University, 20 Garden Street, Cambridge, MA 02138, USA}
\affiliation{Center for Astrophysics $|$ Harvard \& Smithsonian, 60 Garden Street, Cambridge, MA 02138, USA}
\author{Alexandru Lupsasca}
\email{lupsasca@fas.harvard.edu}
\affiliation{Center for the Fundamental Laws of Nature, Harvard University, Cambridge, MA 02138, USA}
\affiliation{Black Hole Initiative at Harvard University, 20 Garden Street, Cambridge, MA 02138, USA}
\affiliation{Society of Fellows, Harvard University, Cambridge, MA 02138, USA}
\author{Andrew Strominger}
\email{strominger@physics.harvard.edu}
\affiliation{Center for the Fundamental Laws of Nature, Harvard University, Cambridge, MA 02138, USA}
\affiliation{Black Hole Initiative at Harvard University, 20 Garden Street, Cambridge, MA 02138, USA}


\begin{abstract}
Black hole images present an annular region of enhanced brightness.  In the absence of propagation effects, this ``photon ring'' has universal features that are completely governed by general relativity and independent of the details of the emission.  Here, we show that the polarimetric image of a black hole also displays universal properties.  In particular, the photon ring exhibits a self-similar pattern of polarization that encodes the black hole spin.  We explore the corresponding universal polarimetric signatures of the photon ring on long interferometric baselines, and propose a method for measuring the black hole spin using a sparse interferometric array.  These signatures could enable spin measurements of the supermassive black hole in M87, as well as precision tests of general relativity in the strong field regime, via a future extension of the Event Horizon Telescope to space.
\end{abstract}

\maketitle

\section{Introduction}

General relativity predicts that a black hole produces a sequence of strongly lensed images of its surrounding emission \cite{Darwin1959,Luminet1979}.  This gravitational lensing results in a striking brightness enhancement (the photon ring) near a critical curve on the screen of a distant observer \cite{Bardeen1973} where these images accumulate (the black hole shadow).  Recent work has revealed a rich and universal substructure for the Kerr photon ring \cite{Johnson2019,Gralla2019b} that is completely governed by general relativity and insensitive to the detailed nature of its astrophysical source.  While the beautiful first image \cite{EHT2019a,EHT2019b,EHT2019c,EHT2019d,EHT2019e,EHT2019f} of M87$^\ast$, the supermassive black hole at the center of the galaxy M87, does not yet resolve the photon ring, a future extension of the Event Horizon Telescope (EHT) to higher frequencies or to space could attain sufficient resolution to observe photon ring substructure \cite{Johnson2019}.

Here, we investigate the \textit{polarimetric} structure of the photon ring.  We show that the photon ring polarization also displays distinctive universal features whose observation could provide a determination of the black hole spin.  More precisely, if the emission region around the black hole is optically thin, then the polarimetric image of the photon ring decomposes into a sequence of subrings, each of which is a lensed image of the direct emission.  The $n^\text{th}$ subring is comprised of photons that circumnavigate the black hole $n/2$ times prior to reaching the telescope.  Subrings with even $n$ or with odd $n$ are found to share the same polarization direction at any given angle around the photon ring, while the change in polarization across consecutive (even and odd) subrings depends on the black hole spin.  In principle, this provides a method for inferring black hole spin from sparse interferometric measurements of the photon ring polarization.

We consider only time-averaged black hole images, which allows us to assume throughout our analysis that the astrophysical sources share (possibly after the time-averaging) the rotational, time-translational and equatorial reflection symmetry of the Kerr geometry itself.\footnote{The more general case involves interesting time-periodic signals, but is beyond the scope of this paper.}  We also assume that propagation effects are negligible, thereby ignoring effects from absorption, scattering, and dispersion.  While we expect absorption and scattering to be weak for EHT observations of M87$^\ast$, whose emission region is thought to be optically thin with an electron density of only $n_{\rm e}\sim10^4\,{\rm cm}^{-3}$ \cite{EHT2019e}, Faraday effects from plasma birefringence are not expected to be negligible \cite{Moscibrodzka2017,Jimenez2018}.  Nevertheless, the universal signature that we describe is comprised of multiple consistency relations and is achromatic, whereas Faraday effects are chromatic.  Thus, frequency-dependent violations of this signature could be used to estimate and remove Faraday effects, as well as to measure properties of the plasma near the black hole.  In addition, all these propagation effects become increasingly weaker at higher observing frequencies.

The paper is organized as follows.  We begin in Sec.~\ref{sec:Imaging} with a review of the propagation of light and its polarization around a Kerr black hole, with an emphasis on the photon shell and ring.  Then, we derive the spin-dependent universal pattern of polarization within the photon ring, which is visible in both its polarimetric image (Sec.~\ref{sec:PhotonRing}) and its interferometric signal (Sec.~\ref{sec:Interferometry}).  We present all our conventions in App.~\ref{app:Polarization} and relegate technical details of the interferometric calculations to App.~\ref{app:Visibility}.

\section{Black hole images \& coordinates}
\label{sec:Imaging}

 In this section, we briefly describe the propagation of light around a Kerr black hole in terms of null geodesics.  We review the photon shell and its almost circular image on the screen of a distant observer: the photon ring.  We present Bardeen coordinates $(\alpha,\beta)$ on the observer screen, and explain their relation to ``ring coordinates'' $(x,y)$ defined relative to the critical curve delineating the ``shadow edge''.  Formulas for the polarization and other details are derived in App.~\ref{app:Polarization}.

\subsection{Light propagation in Kerr}

Photons around an astrophysical black hole of mass $M$ and angular momentum $J=Ma$ propagate along null geodesics in the Kerr geometry, whose metric is given in Boyer-Lindquist coordinates $(t,r,\theta,\phi)$ by
\begin{subequations}
\begin{gather}
	ds^2=-\frac{\Delta}{\Sigma}\pa{\ed t-a\sin^2{\theta}\ed\phi}^2+\frac{\Sigma}{\Delta}\ed r^2+\Sigma\ed\theta^2\nonumber\\
	+\frac{\sin^2{\theta}}{\Sigma}\br{\pa{r^2+a^2}\ed\phi-a\ed t}^2,\\
	\Delta(r)=r^2-2Mr+a^2,\quad 
	\Sigma(r,\theta)=r^2+a^2\cos^2{\theta}.
\end{gather}
\end{subequations}
In Kerr, a photon with affinely-parameterized trajectory $x^\mu(\tau)$ has energy-rescaled four-momentum
\begin{align}
	\label{eq:KerrMomentum}
	p_\mu\ed x^\mu=-\ed t\pm_r\frac{\sqrt{\mathcal{R}(r)}}{\Delta(r)}\ed r\pm_\theta\sqrt{\Theta(\theta)}\ed\theta+\lambda\ed\phi,
\end{align}
given in terms of the radial and angular potentials
\begin{align}
	\mathcal{R}(r)&=\pa{r^2+a^2-a\lambda}^2-\Delta(r)\br{\eta+\pa{\lambda-a}^2},\\
	\Theta(\theta)&=\eta+a^2\cos^2{\theta}-\lambda^2\cot^2{\theta},
\end{align}
where the quantities $(\lambda,\eta)$, which respectively denote the (energy-rescaled) angular momentum parallel to the axis of symmetry and the Carter integral, are conserved along the geodesic.  The photon's trajectory is determined by its initial position $x^\mu(0)$, together with its conserved quantities $(\lambda,\eta)$ and the signs $\pm_r$ and $\pm_\theta$ denoting its initial polar and radial directions of motion.

Let $f^\mu$ denote the linear polarization of the photon, which obeys $f\cdot p=0$ and, in the absence of interactions, the parallel transport law along the photon's trajectory,
\begin{align}
	\label{eq:ParallelTransport}
	p^\mu\nabla_\mu f^\nu=0.
\end{align}
It is a special property of the Kerr spacetime (see, e.g., Refs.~\cite{Chandrasekhar1983,Gates2018}) that, in addition to $\lambda$ and $\eta$, one can define from $f^\mu$ and $p^\mu$ another quantity, the Penrose-Walker constant $\kappa$, which is also conserved along the light ray:
\begin{align}
	\label{eq:PenroseWalker}
	\kappa&=\kappa_1+i\kappa_2
	=\pa{\mathcal{A}-i\mathcal{B}}\pa{r-ia\cos{\theta}},\\
	\mathcal{A}&=\pa{p^tf^r-p^rf^t}+a\sin^2{\theta}\pa{p^rf^\phi-p^\phi f^r},\nonumber\\
	\mathcal{B}&=\br{\pa{r^2+a^2}\pa{p^\phi f^\theta-p^\theta f^\phi}-a\pa{p^tf^\theta-p^\theta f^t}}\sin{\theta}.\nonumber
\end{align}
The conservation of $\kappa$ determines the parallel transport of $f^\mu$ along the null geodesic.  It is the existence of this Penrose-Walker constant $\kappa$ (which ultimately follows from the existence of a conformal Killing tensor on Kerr) that underlies the universality of the polarization pattern described below.

\subsection{Bardeen coordinates and observed polarization}

Bardeen defined Cartesian coordinates $(\alpha,\beta)$ on a distant observer's screen such that the origin $\alpha=\beta=0$ corresponds to a ``line-of-sight'' to the black hole, with the $\beta$-axis the projection of the spin axis onto the plane perpendicular to the line of sight \cite{Bardeen1973}.  In these coordinates, a photon with conserved quantities $(\lambda,\eta)$ that reaches the observer appears on their screen at position
\begin{align}
	\label{eq:BardeenCoordinates}
	\alpha=-\frac{\lambda}{\sin{\theta_\mathrm{o}}},\quad
	\beta=\pm_\mathrm{o}\sqrt{\Theta(\theta_\mathrm{o})},
\end{align}
where $\theta_\mathrm{o}$ denotes the observer's polar inclination from the spin axis and $\pm_\mathrm{o}$ is the sign $\pm_\theta$ of $p^{\theta}$ at the observer.

Given a photon arriving at position $(\alpha,\beta)$, one can retrace its trajectory from the observer back to its source, where its Penrose-Walker constant $\kappa$ can be determined from its initial polarization $f^\mu$ and momentum $p^\mu$.  Its (unit-normalized) observed polarization (direction of the electric field transverse to the photon's momentum) is then computed from $\kappa$ via the relation
\begin{align}
	\label{eq:Polarization}
	\vec{\mathcal{E}}=\pa{\mathcal{E}_\alpha,\mathcal{E}_\beta}
	&=\frac{\pa{\beta\kappa_2-\nu\kappa_1,\beta\kappa_1+\nu\kappa_2}}{\sqrt{\pa{\kappa_1^2+\kappa_2^2}\pa{\beta^2+\nu^2}}},\\
	\label{eq:nu}
	\nu&=-\pa{\alpha+a\sin{\theta_\mathrm{o}}},
\end{align}
The overall sign of $\vec{\mathcal{E}}$ is unphysical, and does not enter the electric vector polarization angle (EVPA)
\begin{align}
	\label{eq:EVPA}
	\chi=\arctan\pa{-\frac{\mathcal{E}_\alpha}{\mathcal{E}_\beta}}.
\end{align}
The EVPA encodes the Stokes parameters $(Q,U)$ as the real and imaginary parts of the complex polarization \cite{Roberts1994}
\begin{align}
	P=Q+iU
	=mIe^{2i\chi},
\end{align}
where $I$ is the Stokes intensity of the ray and $m$ denotes its degree of polarization.  These definitions are consistent with those prevalent in radio astronomy \cite{Hamaker1996}: $\updownarrow$ polarization ($\chi=0$) has $Q>0$, $\leftrightarrow$ polarization ($\chi=\pi/2$) has $Q<0$, $\mathrlap{\searrow}{\nwarrow}$ polarization ($\chi=\pi/4$) has $U>0$, and $\mathrlap{\nearrow}{\swarrow}$ polarization ($\chi=3\pi/4$) has $U<0$ (see also App.~\ref{app:Polarization}).

\subsection{Photon shell and ring}

Light rays with ``critical'' parameters
\begin{align}
	\label{eq:lambdaCritical}
	\tilde{\lambda}&=a+\frac{\tilde{r}}{a}\br{\tilde{r}-\frac{2\tilde{\Delta}}{\tilde{r}-M}},\\
	\label{eq:etaCritical}
	\tilde{\eta}&=\frac{\tilde{r}^{3}}{a^2}\br{\frac{4M\tilde{\Delta}}{\pa{\tilde{r}-M}^2}-\tilde{r}},
\end{align}
asymptote to bound orbits of fixed Boyer-Lindquist radius $\tilde{r}$.  These exist only in the region $\tilde{r}_-\le\tilde r\le\tilde{r}_+$ where 
\begin{align}
	\label{eq:PhotonShell}
	\tilde{r}_\pm=2M\br{1+\cos\pa{\frac{2}{3}\arccos\pa{\pm\frac{a}{M}}}}.
\end{align}
The bound orbits at the boundaries $\tilde{r}=\tilde{r}_\pm$ of this region lie entirely in the equatorial plane, but for other values of $\tilde{r}$, they oscillate in polar angle $\theta$ between
\begin{gather}
	\tilde{\theta}_\pm=\arccos\pa{\mp\sqrt{\tilde{u}_+}},\\
	\tilde{u}_\pm=\tilde{\triangle}\pm\sqrt{\tilde{\triangle}^2+\frac{\tilde{\eta}}{a^2}},\quad
	\tilde{\triangle}=\frac{1}{2}\pa{1-\frac{\tilde{\eta}+\tilde{\lambda}^2}{a^2}}.
\end{gather}
As a bound photon completes a full orbit from $\tilde{\theta}_-$ to $\tilde{\theta}_+$ and back, it also winds in $\phi$ by an azimuth $\delta$ and advances in $t$ by a time lapse $\tau$ computed in Refs.~\citep{Gralla2019b,Teo2003}.

Since every point in the spacetime region 
\begin{align}
	\tilde{r}_-\leq\tilde{r}\leq\tilde{r}_+,\quad
	\tilde{\theta}_-\leq\theta\leq\tilde{\theta}_+,\\
	0\leq\phi<2\pi,\quad
	-\infty\leq t\leq\infty,
\end{align}
has a unique (up to $\pm_\theta$) bound orbit passing through it, we may identify this region of spacetime with the photon shell.  Note that at zero spin, the photon shell contracts to a ``photon sphere'' of radius $\tilde{r}_0=\lim_{a\to0}\tilde{r}_\pm=3M$.

The set of light rays that asymptote to bound orbits in the photon shell defines a closed critical curve on the observer screen parameterized by the orbital radius $\tilde{r}$,
\begin{align}
	\label{eq:CriticalCurve}
	\mathcal{C}&=\pa{\tilde{\alpha}(\tilde{r}),\tilde{\beta}(\tilde{r})},
\end{align}
with $\tilde{\alpha}$ and $\tilde{\beta}$ denoting the Bardeen coordinates \eqref{eq:BardeenCoordinates} evaluated on the critical parameters \eqref{eq:lambdaCritical}--\eqref{eq:etaCritical}.  Note that each radius $\tilde{r}$ gives rise to two points on $\mathcal{C}$ corresponding to the two possible values of the sign $\pm_\mathrm{o}$.  The critical curve $\mathcal{C}$ delineates light rays that terminate on the horizon (inside $\mathcal{C}$) from those that are deflected back to asymptotic infinity (outside $\mathcal{C}$), and coincides with the edge of the ``black hole shadow'' \cite{Bardeen1973,Gralla2019a}.  It also roughly coincides with the location of the photon ring (see Sec.~\ref{sec:PhotonRing}).

Only an equatorial observer sees light from all radii in the photon shell.  Inclined observers still see a closed critical curve arising from asymptotically bound orbits, but these all arise from a subshell $\br{\tilde{r}_\mathrm{min},\tilde{r}_\mathrm{max}}$ of the full photon shell.  The radii $\tilde{r}_\mathrm{min}$ and $\tilde{r}_\mathrm{max}$ are defined by the spin- and observer-inclination-dependent condition $\tilde{\beta}(\tilde{r})=0$.  These radii do not admit a simple closed-form expression but are easy to determine numerically.

\subsection{Ring coordinates and offset}

For almost all values of the black hole spin $a$ and observer inclination $\theta_\mathrm{o}$---excluding only near-extremal black holes ($a\approx M$) viewed by a (nearly) equatorial observer ($\theta_\mathrm{o}\approx\pi/2$)---the critical curve $\mathcal{C}$ is almost perfectly circular \cite{Johannsen2010}.  For example, when $\theta_\mathrm{o}=17^\circ$ (as suggested for M87$^\ast$), the deviation from circularity never exceeds 2\% (see, e.g., Fig.~7 in Ref.~\cite{Johnson2019}).  This makes the experimental prospects for spin measurements from the critical curve's shape challenging, although this may be possible using an extension of the EHT to space \citep{Johnson2019}.  In this work, we instead explore an alternate polarimetric avenue for measuring the spin.

It is natural to describe the black hole image in terms of ``ring coordinates'' $(x,y)$, centered at the origin of the (almost perfect) circle $\mathcal{C}$.  There is a rotational ambiguity in the definition of these coordinates which, if the spin orientation is known (for example, from the brightness variation along the ring \cite{Johnson2019}), can be fixed by demanding that the $y$-axis points along the projected spin axis.  The ring and Bardeen coordinates \eqref{eq:BardeenCoordinates} are then related by a simple translation in $\alpha$ and $x$,
\begin{align}
	\label{eq:RingCoordinates}
	x=\alpha-\Delta\alpha(a,\theta_\mathrm{o}),\quad
	y=\beta,
\end{align}
where\footnote{$\Delta\alpha(a,\theta_\mathrm{o})\approx2a\sin{\theta_\mathrm{o}}$ for all observer inclinations and spins $\ab{a}\lsim\frac{1}{2}$.}
\begin{align}
	\Delta\alpha=\frac{\tilde{\alpha}(\tilde{r}_\mathrm{max})+\tilde{\alpha}(\tilde{r}_\mathrm{min})}{2}.
\end{align}
For an equatorial observer seeing the entire photon shell, $\tilde{r}_\mathrm{max/min}=\tilde{r}_\pm$, and it is possible to obtain an analytic expression for the shift $\Delta\alpha$.  For other observer inclinations $\theta_\mathrm{o}\neq\pi/2$, this shift depends on the numerically-determined values of $\tilde{r}_\mathrm{max/min}$.

\section{Photon ring polarization}
\label{sec:PhotonRing}

This section extends to the polarization Stokes parameters $Q$, $U$, and $V$ the analysis of the Kerr photon ring substructure presented in Refs.~\cite{Johnson2019,Gralla2019b} for the intensity $I$.  As with image intensity, we find that the polarimetric image of the photon ring also displays universal features, which we explore in detail.  In particular, we show that this universal pattern of polarization encodes the black hole spin, and may in principle provide a new method of spin measurement.

\subsection{Universal polarimetric image of the photon ring}

\begin{figure}
	\centering
	\includegraphics[width=\columnwidth]{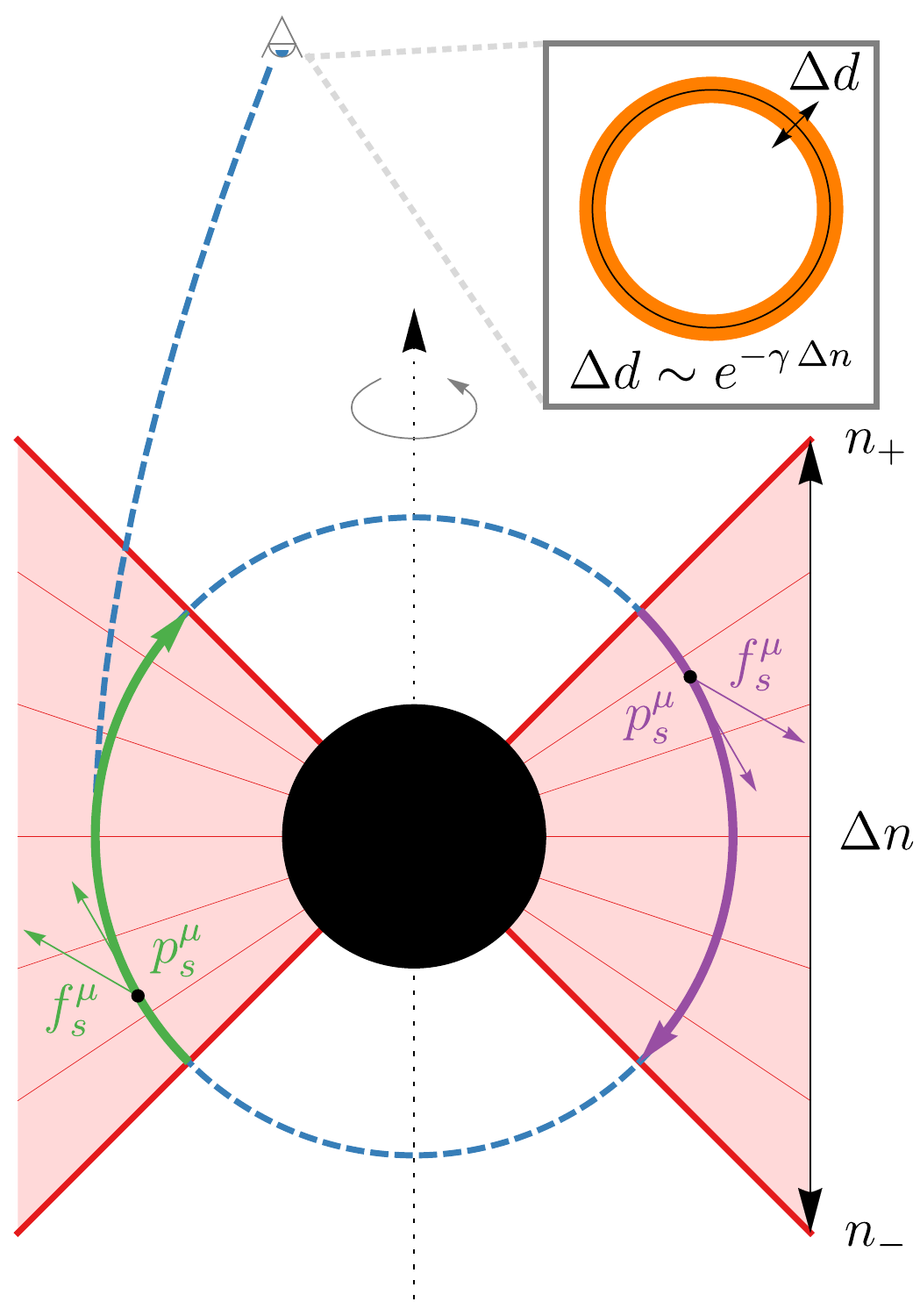}
	\caption{Ray-tracing into a geometrically thick, optically thin bulk matter distribution.  Light rays near a point $(\tilde{\alpha}(\tilde{r}),\tilde{\beta}(\tilde{r}))$ on the critical curve are nearly bound at the photon shell radius $r=\tilde{r}$ and can complete multiple orbits through the emission region.  Even/odd half-orbits travel through the same matter distribution at the same inclination, and hence are loaded with photons making identical contributions to the Stokes parameters of the ray.  Light rays executing $n\in\br{n_-,n_+}$ half-orbits appear within a narrow window $\Delta d\sim e^{-\gamma(n_+-n_-)}$ near the critical curve on the screen.}
	\label{fig:BulkMatter}
\end{figure}

In radiative transport, the optical appearance of matter emissions is computed by ray-tracing null geodesics backwards from the observer into the source, integrating the contributions of photons collected or absorbed along the ray.  Light rays shot backwards from a point lying at a small perpendicular distance $d$ from the critical curve $\mathcal{C}$ on the observer screen complete a logarithmically divergent (fractional) number of half-orbits as $d\to0$ and the point approaches the critical curve \cite{Johnson2019,Gralla2019b},\footnote{The fractional orbit number is defined precisely in Eq.~(36) of Ref.~\cite{Gralla2019b}, where Eq.~\eqref{eq:nDivergence} also appears as Eq.~(74) with two differences: 1) an extra factor of 2 accounting for the definition of $n$ therein as the number of orbits rather than half-orbits, and 2) an additional subleading coefficient $\hat{C}_\pm$ that improves the logarithmic approximation and is necessary for it to hold near $n\approx1$.}
\begin{align}
	\label{eq:nDivergence}
	n\sim-\frac{1}{\gamma}\log\ab{d},
\end{align}
where the Lyapunov exponent $\gamma(\tilde{r})$, which characterizes the instability of the orbit at radius $\tilde{r}$, is given by
\begin{align}
	\gamma=\frac{4}{a}\sqrt{\tilde{r}^2-\frac{M\tilde{r}\Delta(\tilde{r})}{\pa{\tilde{r}-M}^2}}\int_0^1\!\frac{\ed t}{\sqrt{\pa{1-t^2}\pa{\tilde{u}_+t^2-\tilde{u}_-}}}.
\end{align}

If the bulk matter distribution is geometrically thick with a gap near the poles (as depicted in Fig.~\ref{fig:BulkMatter}), then an additional passage through the matter region extending from fractional orbit number $n_-$ to $n_+$ requires the light ray to be aimed within an exponentially narrow window on the observer screen, of width
\begin{align}
	\Delta d=d_+-d_-
	\sim e^{-\gamma\pa{n_+-n_-}}.
\end{align}
Successive images of the emitting region are labeled by the half-orbit number $n$.  In particular, they differ by an integer shift $n_\pm\to n_\pm+1$, corresponding to an exponential demagnification $d_\pm\to d_\pm e^{-\gamma}$ on the observer screen.  As such, an optically transparent bulk matter distribution produces an infinite self-similar sequence of exponentially demagnified subrings, which are labeled by half-orbit number $n$ and have exponentially narrow width
\begin{align}
	\label{eq:OnAxisDemagnification}
	\Delta d_{n+1}=e^{-\gamma}\Delta d_n.
\end{align}
Note that the radial dependence of the orbital instability in the photon shell introduces an angle-dependence of the demagnification factor $e^{-\gamma(\tilde{r})}$ around the photon ring.

Light rays that execute $n$ half-orbits around the black hole are exponentially close in $n$ to their nearby bound orbit, i.e., their arrival position $(\alpha_n,\beta_n)=(\tilde{r},d_n)$ on the observer screen converges exponentially in $n$ to the nearest point $(\alpha(\tilde{r}),\beta(\tilde{r}))$ on the critical curve corresponding to their nearby bound orbit.  Since $d_n\to0$ exponentially in $n$, the conserved quantities $(\lambda_n,\eta_n)$ also converge to their values $(\tilde{\lambda}(\tilde{r}),\tilde{\eta}(\tilde{r}))$ on the nearby bound orbit exponentially in $n$.  As such, all the photon momenta $p^\mu(\lambda_n,\eta_n)$ along the portion of these rays inside the photon shell are approximately equal to their limiting value $\tilde{p}^\mu=p^\mu(\tilde{\lambda},\tilde{\eta})$.  Note however that the sign of $p^\theta$ during the last passage in the photon shell depends on the parity of $n$ (green/purple arrows in Fig.~\ref{fig:BulkMatter}).

Taking all this into account, it follows that the light rays corresponding to the $n^\text{th}$ and $(n+2)^\text{th}$ subrings sweep through the matter distribution with almost the same momentum $\tilde{p}^\mu$, up to corrections that are exponentially suppressed in $n$ and already negligible for $n\gsim1$.  In particular, they travel through the emitting matter at essentially the same inclination, and hence are loaded with the same distribution of Stokes parameters $\vec{S}=\pa{I,Q,U,V}$ during each even/odd pass through the emitting region:
\begin{align}
	\label{eq:Subrings}
	\vec{S}_{n+2}^\mathrm{ring}\pa{d_{n+2}}=\vec{S}_n^\mathrm{ring}\pa{d_n}.
\end{align}

In the case where the photons are emitted isotropically, the sign of $p^\theta$ during the last passage through the emission region is irrelevant.  Therefore, the $n^\text{th}$ and $(n+1)^\text{th}$ subrings pick up the same contribution to the intensity,
\begin{align}
	\label{eq:OnAxisSubringIntensity}
	I_{n+1}^\mathrm{ring}\pa{d_{n+1}}=I_n^\mathrm{ring}\pa{d_n}.
\end{align}
Together, these subrings give rise to a logarithmic brightness enhancement near the critical curve $\mathcal{C}$: this is the origin of photon ring.

\begin{figure}
	\centering
	\includegraphics[width=\columnwidth]{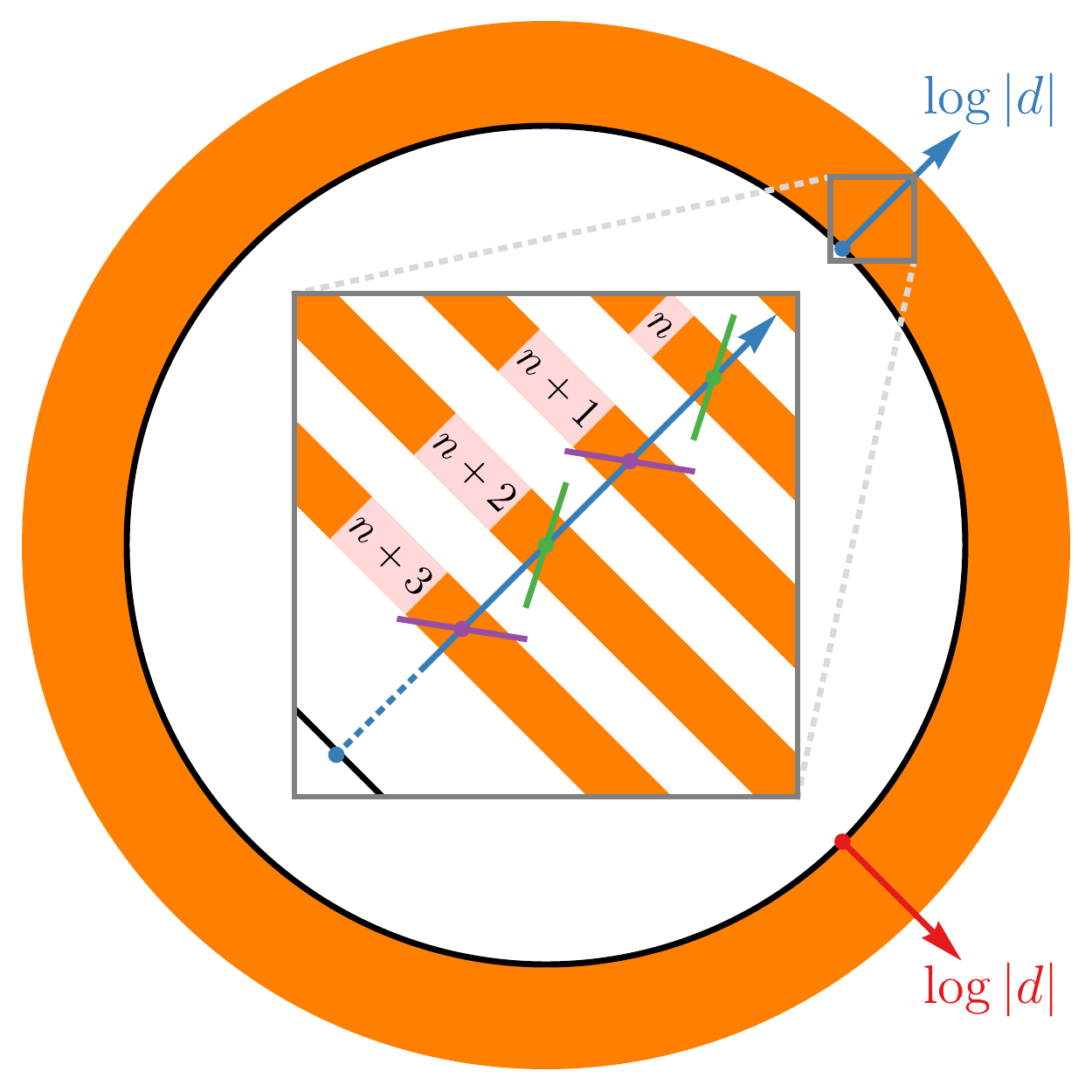}
	\caption{The photon ring and its universal substructure: a self-similar sequence of subrings labeled by half-orbit number $n$ and exponentially converging to the critical curve (black).  The subrings are exponentially demagnified images of the main emission appearing at perpendicular distance from the critical curve $d_{n+1}\sim e^{-\gamma}d_n$.  Hence, on a logarithmic scale, the subrings have equal width and equal spacing.  Even/odd subrings display the same polarization $\vec{\mathcal{E}}$, depicted with green/purple ticks and corresponding to the contributions collected on the green/purple passes through the matter region depicted in Fig.~\ref{fig:BulkMatter}.  Importantly, the polarization alternates between even and odd subrings in a manner that reflects the black hole spin.}
	\label{fig:PhotonRing}
\end{figure}

According to Eq.~\eqref{eq:Subrings}, the Stokes parameters $Q$, $U$, and $V$ exhibit the same ``wedding-cake layer'' structure as the intensity $I$ described in Ref.~\cite{Johnson2019}, except that the layers may now alternate in height according to the parity of $n$, and in particular may be negative since these parameters are not necessarily positive, unlike $I$.

Moreover, the linear polarization encoded in $Q$ and $U$ has a simple visual description: Eqs.~\eqref{eq:Polarization} and \eqref{eq:Subrings} together imply that the subring polarization ticks obey
\begin{align}
	\vec{\mathcal{E}}_{n+2}=\vec{\mathcal{E}}_n.
\end{align}
This universal pattern is illustrated in Fig.~\ref{fig:PhotonRing}.\footnote{The gap between subrings in Fig.~\ref{fig:PhotonRing} occurs when the emission region is geometrically thick with a gap near the poles and lies outside the photon shell (in which case the photon ring lies strictly outside the critical curve and its subrings do not overlap).}

In general, the subring intensities obey Eq.~\eqref{eq:Subrings}, and so $I_{n+2}^\mathrm{ring}\pa{d_{n+2}}=I_n^\mathrm{ring}\pa{d_n}$.  As we argued above Eq.~\eqref{eq:OnAxisSubringIntensity}, this relation between even and odd subrings can be extended to all consecutive subrings whenever photons are emitted isotropically, so that $I_{n+1}^\mathrm{ring}\pa{d_{n+1}}=I_n^\mathrm{ring}\pa{d_n}$.  A similar extension is not possible for the polarization Stokes parameters $Q$, $U$, and $V$, since their emission is fundamentally directed and therefore cannot be isotropic.  For instance, in models of synchrotron radiation emission, the source polarization $f^\mu$ depends on the local orientation of the electromagnetic field $F_{\mu\nu}$.  Nonetheless, it is possible to connect the polarization across consecutive subrings for bulk matter distributions satisfying stationarity, axisymmetry, $and$ equatorial reflection symmetry.  Realistic matter distributions seem likely to satisfy these conditions after time-averaging, so we expect our conclusions to hold generally for time-averaged images.

Given this time-averaged symmetry, for every photon loaded at polar angle $\theta_\mathrm{s}$ onto a ray executing $n$ passes through the emission region, there is a corresponding photon loaded at polar angle $\pi-\theta_\mathrm{s}$ onto the nearby ray executing $n+1$ passes, such that
\begin{align}
	\label{eq:MomentumFlip}
	\br{p_{n+1}^\theta}_{\pi-\theta_\mathrm{s}}&\approx-\br{p_n^\theta}_{\theta_\mathrm{s}},\\
	\label{eq:PolarizationFlip}
	\br{f_{n+1}^\theta}_{\pi -\theta_\mathrm{s}}&\approx-\br{f_n^\theta}_{\theta_\mathrm{s}},
\end{align}
with all other components of $p_\mathrm{s}^\mu$ and $f_\mathrm{s}^\mu$ equal (see Fig.~\ref{fig:BulkMatter}).

It is manifest from its definition \eqref{eq:PenroseWalker} that the Penrose-Walker constant transforms as $\kappa\to\bar{\kappa}$ when $p^\theta\to-p^\theta$, $f^\theta\to-f^\theta$, $\theta\to\pi-\theta$, with all other quantities kept equal (since $\mathcal{A}$ remains invariant while $\mathcal{B}$ changes sign).  Hence, comparing the Penrose-Walker constant of a photon loaded onto the $n^\text{th}$ ray at $\theta_\mathrm{s}$ with that of a photon loaded onto the $(n+1)^\text{th}$ ray at the reflected angle $\pi-\theta_\mathrm{s}$, Eqs.~\eqref{eq:MomentumFlip} and \eqref{eq:PolarizationFlip} imply that
\begin{align}
	\br{\kappa_{n+1}}_{\pi-\theta_\mathrm{s}}&=\br{\bar{\kappa}_n}_{\theta_\mathrm{s}}.
\end{align}
Integrating over the entire polar region of emission, the net polarizations loaded onto each ray obey
\begin{align}
	\label{eq:SubringsKappa}
	\kappa_{n+1}^\mathrm{ring}\pa{d_{n+1}}=\bar{\kappa}_n^\mathrm{ring}\pa{d_n}.
\end{align}
Note that, as the sign of $\kappa$ is unphysical, there is no rotation of the polarization in the (unnatural) special case that $\kappa$ is either purely real or purely imaginary. 

Reflection symmetry about the equatorial plane also implies that the image is reflection-symmetric about the horizontal axis on the observer screen.  Every photon shell radius $\tilde{r}$ visible to a distant observer produces two images on the critical curve $\mathcal{C}$ on the observer screen, which differ by $\tilde{\beta}\to-\tilde{\beta}$ and arise from rays hitting the observer from above or below [the sign $\pm_\mathrm{o}$ in Eq.~\eqref{eq:BardeenCoordinates}].  These rays' last pass through their corresponding photon shell radius $\tilde{r}$ is executed in opposite directions, differing by $p^\theta\to-p^\theta$.  Hence, by the same argument given above, one ray is loaded with Penrose-Walker constant $\kappa$ and the other with $\bar{\kappa}$, so that
\begin{align}
	\label{eq:SubringSymmetry}
	\kappa_{n}^\mathrm{ring}\pa{\beta}=\bar{\kappa}_n^\mathrm{ring}\pa{-\beta}.
\end{align}

\subsection{Black hole spin from photon ring polarization}

According to Eq.~\eqref{eq:SubringsKappa}, for every photon with Penrose-Walker constant $\kappa$ loaded onto a ray passing $n$ times through a reflection-symmetric emission region, there is a corresponding photon with Penrose-Walker constant $\bar{\kappa}$ loaded onto a nearby ray passing $n+1$ times through the same emission region.  As such, sums and differences of the polarization ticks across successive subrings (the green and purple ticks in Fig.~\ref{fig:PhotonRing}) obey
\begin{align} 
	\vec{\mathcal{E}}_+&=\frac{1}{2}\pa{\vec{\mathcal{E}}_{n+1}+\vec{\mathcal{E}}_n}
	=\frac{(-\tilde{\nu}\kappa_1,\tilde{\beta}\kappa_1)}{\sqrt{\pa{\kappa_1^2+\kappa_2^2}(\tilde{\beta}^2+\tilde{\nu}^2)}},\\
	\vec{\mathcal{E}}_-&=\frac{1}{2}\pa{\vec{\mathcal{E}}_{n+1}-\vec{\mathcal{E}}_n}
	=\frac{(-\tilde{\beta}\kappa_2,-\tilde{\nu}\kappa_2)}{\sqrt{\pa{\kappa_1^2+\kappa_2^2}(\tilde{\beta}^2+\tilde{\nu}^2)}},
\end{align}
where $\kappa_1+i\kappa_2=\kappa_n$ depends entirely on the nature of the astrophysical source, while the repeated pattern encoded in $\vec{\mathcal{E}}_\pm$ is a universal feature of general relativity.  Note that certain (orientation-independent) ratios of the components of $\vec{\mathcal{E}}_\pm$ give universal functions of $\tilde{r}$ and $a$:
\begin{align}
	\frac{\mathcal{E}_+^\alpha}{\mathcal{E}_+^\beta}=-\frac{\mathcal{E}_-^\beta}{\mathcal{E}_-^\alpha}
	=\frac{\tilde{\alpha}+a\sin{\theta_\mathrm{o}}}{\tilde{\beta}}.
\end{align}
This relation provides an observational prospect for measuring the black hole spin $a$ from the universal pattern of polarization in the photon ring (as well as a consistency check that can be used to confirm that the observed physics is indeed in the universal regime):
\begin{align}
	a\sin{\theta_\mathrm{o}}=\tilde{\beta}\frac{\mathcal{E}_+^\alpha}{\mathcal{E}_+^\beta}-\tilde{\alpha}
	=-\tilde{\beta}\frac{\mathcal{E}_-^\beta}{\mathcal{E}_-^\alpha}-\tilde{\alpha}.
\end{align}
Conversely, a violation of this relation would imply that effects such as Faraday rotation are not in fact negligible.

Finally, this pattern can also be described in terms of the EVPA \eqref{eq:EVPA}: at fixed angle $\tilde{r}$ around the photon ring,
\begin{align}
	\tan\pa{\chi_n+\chi_{n+1}}=\frac{2\tilde{\beta}\pa{\tilde{\alpha}+a\sin{\theta_\mathrm{o}}}}{\pa{\tilde{\alpha}+a\sin{\theta_\mathrm{o}}}^2-\tilde{\beta}^2}.
\end{align}

\subsection{Consistency relations from image symmetry}

According to Eq.~\eqref{eq:SubringSymmetry}, every subring has a reflection-symmetric polarimetric image.  By Eqs.~\eqref{eq:Polarization} and \eqref{eq:EVPA},
\begin{align}
	\chi=\arctan\pa{\frac{\nu\kappa_1-\beta\kappa_2}{\beta\kappa_1+\nu\kappa_2}}.
\end{align}
Thus, if the ray at position $\beta$ on the $n^\text{th}$ subring has Penrose-Walker constant $\kappa$ with corresponding EVPA $\chi$, then by Eq.~\eqref{eq:SubringSymmetry}, the ray at reflected position $-\beta$ on the same subring has Penrose-Walker constant $\bar{\kappa}$ with corresponding EVPA $-\chi$.  Hence,
\begin{align}
	\chi_n\pa{\beta}+\chi_n\pa{-\beta}=0.
\end{align}
This relation may be viewed as a test for the assumptions behind its derivation: its violation would imply that the emission region is not reflection-symmetric, or else that its influence on the propagation of light is nonnegligible.

\section{Interferometric signatures}
\label{sec:Interferometry}

In this section, we explore the universal signatures of the photon ring on long interferometric baselines.  After reviewing the image visibility $\hat{I}$ of the photon ring \cite{Johnson2019}, we derive the universal pattern encoded in its Fourier-transformed complex polarization $\hat{P}$.

\subsection{Visibility functions for thin rings}

\begin{figure*}
	\centering
	\includegraphics[width=\textwidth]{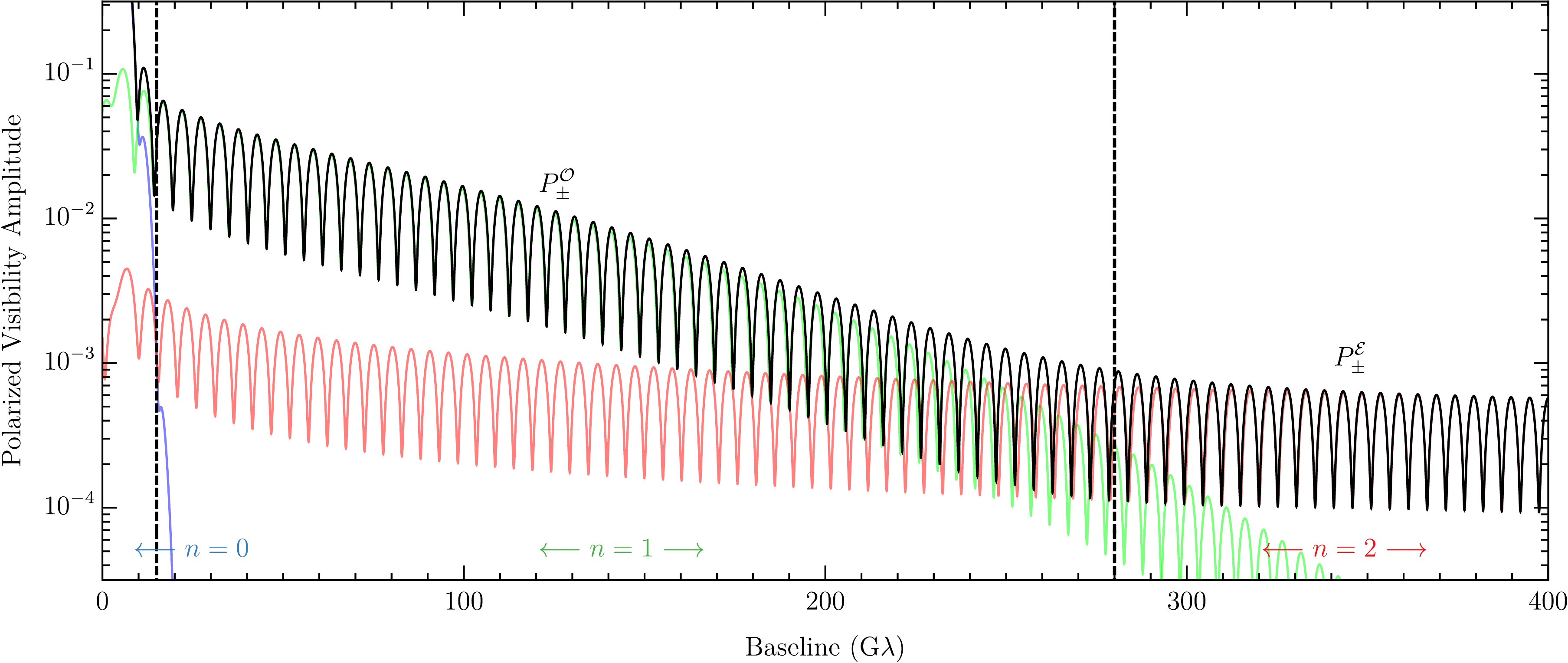}
	\caption{Schematic showing polarized visibility amplitude as a function of baseline length for a photon ring with $d=40\,\mu{\rm as}$. Visibility amplitudes are shown for the complete image (black) as well as for individual subrings ($n=0$: blue, 1: green, 2: red). 
	Progressively longer baselines are dominated by visibilities for a subring with correspondingly larger index $n$. The visibility amplitudes in each range are determined by a pair of complex coefficients, $P_\pm^{\mathcal{O}}$ or $P_\pm^{\mathcal{E}}$ depending upon whether $n$ is odd or even. The periodicity of the visibility amplitudes gives the subring diameter in the baseline direction, while the coefficients $P_\pm^{\mathcal{O}}$ and $P_\pm^{\mathcal{E}}$ carry information about the black hole spin and azimuthal brightness asymmetry.}
	\label{fig:VisibilityCartoon}
\end{figure*}

Interferometers observe an image of an astrophysical source $I(\mathbf{x})$ by sampling baselines of its Fourier transform
\begin{align}
	\hat{I}(\mathbf{u})=\int I(\mathbf{x})e^{-2\pi i\mathbf{u}\cdot\mathbf{x}}\ed^2\mathbf{x},
\end{align}
where $\mathbf{u}$ is the baseline vector in units of observation wavelength and $\mathbf{x}=(-x,y)$ is the image coordinate in radians (these coordinates match astronomical image conventions).  $\hat{I}(\mathbf{u})$ is called the complex visibility.  Likewise, the complex polarization $P(\mathbf{x})$ is measured by sampling
\begin{align}
	\hat{P}(\mathbf{u})=\int P(\mathbf{x})e^{-2\pi i\mathbf{u}\cdot\mathbf{x}}\ed^2\mathbf{x}.
\end{align}
As in Ref.~\cite{Johnson2019}, we use polar coordinates $(u,\varphi_u)$ in the baseline plane and $(\rho,\varphi_\rho)$ on the observer screen, where
\begin{align}
	\label{eq:PolarCoordinates}
	\rho=\sqrt{x^2+y^2},\quad
	\varphi_\rho=\arctan\pa{-\frac{x}{y}},
\end{align}
such that $y-ix=\rho e^{i\varphi_\rho}$, with the angle $\varphi_\rho$ increasing counterclockwise from the vertical (the spin axis).

A function $F(\rho,\varphi_\rho)$ with support localized on a thin, circular ring of diameter $d$ on the observer screen decomposes into a sum over angular Fourier modes:
\begin{align}
	\label{eq:ThinRing}
	F(\rho,\varphi_\rho)=\frac{1}{\pi d}\delta\pa{\rho-\frac{d}{2}}\sum_{m=-\infty}^\infty F_me^{im\varphi_\rho}.
\end{align}
The corresponding visibility function is then \cite{Johnson2019}
\begin{align}
	\label{eq:RingVisibility}
	\hat{F}(u,\varphi_u)&=\sum_{m=-\infty}^\infty F_mJ_m(\pi du)e^{im\pa{\varphi_u-\pi/2}}\\
	&\approx\frac{F_+(\varphi_u)\cos\pa{\pi du}+F_-(\varphi_u)\sin\pa{\pi du}}{\sqrt{du}},\nonumber\\
	F_\pm(\varphi_u)&=\frac{1}{\pi}\sum_{m=-\infty}^\infty F_me^{im\br{\varphi_u+\frac{\pi}{2}\pa{m-1\pm1}}}.
\end{align}
Here, $J_m$ denotes the $m^\text{th}$ Bessel function of the first kind, and its asymptotic expansion is a valid approximation on long baselines $u\gg m_\mathrm{max}^2/(\pi d)$.  In principle, an interferometer with baselines that sample all angles $\varphi_u$ can measure the full set of coefficients $\cu{F_m}$, while a single-baseline interferometer at a fixed angle $\varphi_u$ is only sensitive to $F_\pm(\varphi_u)$.  Both configurations can measure the ring diameter $d$ via the radial baseline dependence.

Finally, note that if the image function $F$ is complex-conjugated under reflections about the horizontal axis, i.e., $F(\alpha,-\beta)=\overline{F(\alpha,\beta)}$, then
\begin{align}
	F(\rho,\varphi_\rho)=\overline{F(\rho,\pi-\varphi_\rho)}.
\end{align}
Together with Eq.~\eqref{eq:ThinRing}, it then follows that
\begin{align}
	\label{eq:ImageSymmetry}
	\bar{F}_m=(-1)^mF_m.
\end{align}

\subsection{Complex visibility of the photon ring}

As reviewed in Sec.~\ref{sec:PhotonRing}, the photon ring is composed of subrings that exponentially converge to the critical curve $\mathcal{C}$, which is an almost perfect circle.  Letting $d$ denote the diameter of $\mathcal{C}$, it follows that the $n^\text{th}$ subring is a thin annulus of width $w_n=e^{-\gamma}w_{n-1}$ and diameter $d\gg w_n$.  (This holds for $n\gsim1$, with corrections to the diameter exponentially suppressed in the half-orbit number $n$.)

For the $n^\text{th}$ subring, the delta-function approximation \eqref{eq:ThinRing} is valid on baselines $u$ that are long enough to resolve the diameter of the ring, but not its width:
\begin{align}
	\frac{1}{d}\ll u\ll\frac{1}{w_n}.
\end{align}
On longer baselines $u\gg1/w_n$, the annulus is resolved and its visibility decays faster than the $1/\sqrt{u}$ fall-off \eqref{eq:RingVisibility}.  Hence, the $n^\text{th}$ subring dominates the signal in the regime
\begin{align}
	\label{eq:SubringRegimes}
	\frac{1}{w_{n-1}}
	\ll u
	\ll\frac{1}{w_n},
\end{align}
in which its complex visibility is [Eq.~\eqref{eq:RingVisibility} with $F=I$]
\begin{align}
	\label{eq:ImageVisibility}
	\hat{I}_n^\mathrm{ring}(u)\approx\frac{w_n}{\sqrt{u}}.
\end{align}
As such, the photon ring has complex visibility
\begin{align}
	\hat{I}^\mathrm{ring}(u)\approx\sum_{\substack{n\\w_n<1/u}}\hat{I}_n^\mathrm{ring}(u)
	\sim\frac{1}{u^{3/2}},
\end{align}
which forms a cascade of damped oscillations on progressively longer baselines (see Ref.~\cite{Johnson2019} and Fig.~5 therein).

\subsection{Polarization visibility of the photon ring}

\begin{figure}
	\centering
	\includegraphics[width=\columnwidth]{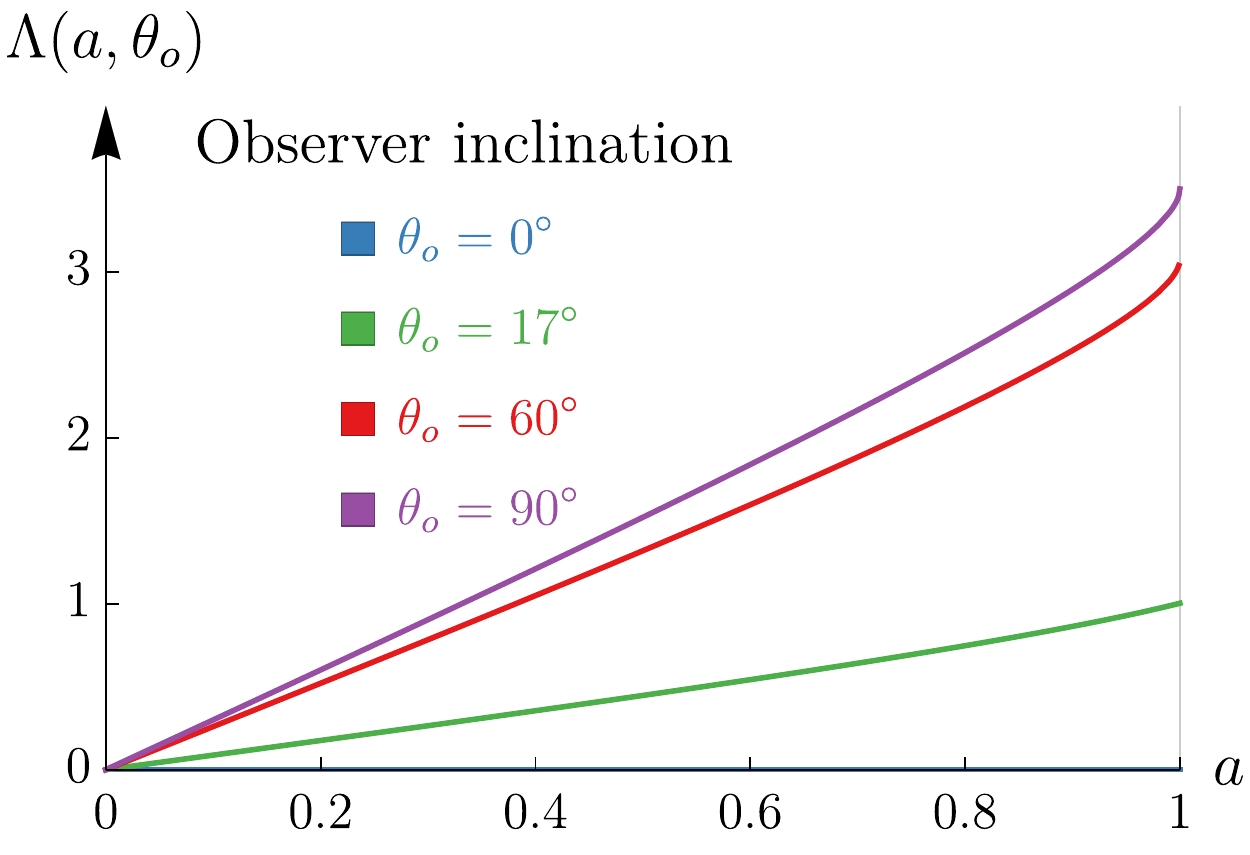}
	\caption{The offset function $\Lambda(a,\theta_\mathrm{o})$ is a bijective function of the black hole spin parameter $a$.  We set $M=1$ in this plot.}
	\label{fig:Lambda}
\end{figure}

By Eq.~\eqref{eq:Subrings}, all even subrings share the same complex polarization, as do all the odd subrings:
\begin{align}
	P_{n+2}^\mathrm{ring}(d_{n+2},\varphi_\rho)=P_n^\mathrm{ring}(d_n,\varphi_\rho).
\end{align}
In Fourier space, the $n^\text{th}$ subring dominates the signal in the regime \eqref{eq:SubringRegimes}, in which the polarization visibility scales like the complex visibility \eqref{eq:ImageVisibility} for the image intensity,
\begin{align}
	\hat{P}_n^\mathrm{ring}(u)\sim\frac{w_n}{\sqrt{u}}.
\end{align}
Fig.~\ref{fig:VisibilityCartoon} depicts this universal cascading pattern of polarization on long interferometric baselines.

Further constraints emerge when considering time-averaged (reflection-symmetric) emissions.  First, note that the image-plane complex polarization $P(\rho,\varphi_\rho)$ naturally decomposes into [Eq.~\eqref{eq:PolarizationPW} below]
\begin{align}
	P=\pa{\beta+i\nu}^2\mathcal{P},\quad
	\mathcal{P}=\pa{\frac{mI}{\beta^2+\nu^2}}\frac{\bar{\kappa}}{\kappa},
\end{align}
where $\nu$ here is the spin-dependent ring offset defined in Eq.~\eqref{eq:nu}.  By the preceding discussion [Eq.~\eqref{eq:SubringsKappa} above], $\mathcal{P}(\rho,\varphi_\rho)$ is therefore complex-conjugated across subrings,
\begin{align}
	\label{eq:SubringRelation}
	\mathcal{P}_{n+1}^\mathrm{ring}\pa{d_{n+1},\varphi_\rho}=\bar{\mathcal{P}}_n^\mathrm{ring}\pa{d_n,\varphi_\rho}.
\end{align}
Thus, even and odd subrings have complex polarizations
\begin{align}
	\label{eq:RelationToP}
	P^\mathcal{E}=\pa{\beta+i\nu}^2\bar{\mathcal{P}},\quad
	P^\mathcal{O}=\pa{\beta+i\nu}^2\mathcal{P}.
\end{align}
Eliminating $\mathcal{P}$, one obtains the fundamental relation
\begin{align}
	\label{eq:EvenOddSubrings}
	\pa{\beta+i\nu}^2\overline{P^\mathcal{E}}=\pa{\beta-i\nu}^2P^\mathcal{O}.
\end{align}
In addition, $\mathcal{P}(\rho,\varphi_\rho)$ is also complex-conjugated under reflections about the horizontal axis within each subring [Eq.~\eqref{eq:SubringSymmetry} above].  Hence, under image reflections $\beta\to-\beta$,
\begin{align}
	\label{eq:SubringConjugation}
	P^{\mathcal{E},\mathcal{O}}\pa{\alpha,-\beta}=\overline{P^{\mathcal{E},\mathcal{O}}\pa{\alpha,\beta}}.
\end{align}

Using the notation introduced in Eqs.~\eqref{eq:ThinRing} and \eqref{eq:RingVisibility}, the image-plane functions
\begin{align}
	\label{eq:PolarizationExpansion}
	P^{\mathcal{E},\mathcal{O}}(\rho,\varphi_\rho)=\frac{1}{\pi d}\delta\pa{\rho-\frac{d}{2}}\sum_{m=-\infty}^\infty P_m^{\mathcal{E},\mathcal{O}}e^{im\varphi_\rho}
\end{align}
have Fourier coefficients $P_m^{\mathcal{E},\mathcal{O}}$ and corresponding visibility functions $\hat{P}^{\mathcal{E},\mathcal{O}}$.  By Eqs.~\eqref{eq:ImageSymmetry} and \eqref{eq:SubringConjugation}, these Fourier coefficients obey
\begin{align}
    \label{eq:ConjugateCoefficients}
	\overline{P_m^{\mathcal{E},\mathcal{O}}}=(-1)^mP_m^{\mathcal{E},\mathcal{O}}.
\end{align}
This condition provides a useful way of testing that the reflection axis of the image has been correctly identified.

\subsection{Estimating black hole spin}

Finally, we describe two methods for estimating black hole spin from the polarization of successive subrings.

Since $\beta+i\nu=\beta-i\pa{\alpha+a\sin{\theta_\mathrm{o}}}$, it follows from Eqs.~\eqref{eq:RingCoordinates}, \eqref{eq:Offset}, and \eqref{eq:PolarCoordinates} that
\begin{align}
	\label{eq:BetaUpsilon}
	\beta+i\nu=\rho e^{i\varphi_\rho}-i\Lambda(a,\theta_\mathrm{o}),
\end{align}
where we introduced the `offset' function
\begin{align}
	\label{eq:Offset}
	\Lambda(a,\theta_\mathrm{o})=\Delta\alpha(a,\theta_\mathrm{o})+a\sin{\theta_\mathrm{o}},
\end{align}
which is, conveniently, both a directly observable quantity (as we will now show) and a monotonically increasing function of black hole spin $a$ (see Fig.~\ref{fig:Lambda}).  Measuring $\Lambda$ at a given $\theta_\mathrm{o}$ is hence equivalent to measuring the spin.

Combining this with Eqs.~\eqref{eq:EvenOddSubrings} and \eqref{eq:PolarizationExpansion}, which together imply that
\begin{align}
	\pa{\beta+i\nu}^2\sum_{m=-\infty}^\infty\overline{P_m^\mathcal{E}}e^{-im\varphi_\rho}=\pa{\beta-i\nu}^2\sum_{m=-\infty}^\infty P_m^\mathcal{O}e^{im\varphi_\rho},\nonumber
\end{align}
we find that near the critical curve $\rho=d/2$,
\begin{align}
	\frac{d^2}{4}\pa{P_{m+2}^\mathcal{O}-\overline{P_{-m+2}^\mathcal{E}}}&+id\Lambda\pa{P_{m+1}^\mathcal{O}+\overline{P_{-m+1}^\mathcal{E}}}\nonumber\\
	&-\Lambda^2\pa{P_m^\mathcal{O}-\overline{P_{-m}^\mathcal{E}}}=0.
\end{align}
In terms of real coefficients defined as\footnote{It follows from Eq.~\eqref{eq:ConjugateCoefficients} that the coefficients $C_{m,k}$ transform as $\overline{C_{m,k}}=C_{m,k}$ under complex conjugation and as $C_{m,k}\to-C_{-m,k}$ under $\mathcal{E}\leftrightarrow\mathcal{O}$ interchange.  Either way, Eq.~\eqref{eq:SpinCoefficients} remains invariant.}
\begin{align}
	\label{eq:Coefficients}
	C_{m,k}=i^{m+k}\br{P_{m+k}^\mathcal{O}-(-1)^k\overline{P_{-m+k}^\mathcal{E}}}\in\mathbb{R},
\end{align}
this can be more simply rewritten as
\begin{align}
	\label{eq:SpinCoefficients}
	\pa{\frac{\Lambda}{d}}^2C_{m,0}-\pa{\frac{\Lambda}{d}}C_{m,1}+\frac{C_{m,2}}{4}=0.
\end{align}
Since $d$ is easily measured from the radial periodicity alone, one can thus estimate $\Lambda$ from the coefficients \eqref{eq:Coefficients}, and hence infer the black hole spin from the relation\footnote{Plugging in Eq.~\eqref{eq:CoefficientRelations} reveals that the quantity under the square root is itself the square of a real quantity, and therefore positive.  Hence, there are always two real roots for $\Lambda$; if they are both physically admissible, then the spin is only determined up to a two-fold degeneracy, which could be resolved by varying $m$.}
\begin{align}
	\label{eq:SpinSolution}
	\Lambda=\frac{d}{2}\pa{\frac{C_{m,1}}{C_{m,0}}\pm\sqrt{\frac{C_{m,1}^2}{C_{m,0}^2}-\frac{C_{m,2}}{C_{m,0}}}}.
\end{align}
The modes $P_\pm^{\mathcal{O}}(\varphi_u)$ and $P_\pm^{\mathcal{E}}(\varphi_u)$ measurable with individual baselines also satisfy some elegant relationships.  For a specific baseline angle $\varphi_u$, we define the quantities
\begin{align}
	\label{eq:Z}
	Z_k^\pm(\varphi_u)&=\br{P_+^\mathcal{O}(\varphi_u)\pm(-1)^kP_-^\mathcal{O}(\varphi_u)}e^{-ik\varphi_u}\nonumber\\
	&\quad \mp \br{\overline{P_+^\mathcal{E}(\varphi_u)}\pm(-1)^k\overline{P_-^\mathcal{E}(\varphi_u)}}e^{ik\varphi_u}.
\end{align}
We show in App.~\ref{app:Visibility} that spin may be inferred from measurements on a single radial baseline $\varphi_u$ via\footnote{As with Eq.~\eqref{eq:SpinSolution}, if both roots are physical, then the spin is only determined up to a two-fold degeneracy, which may be resolved by varying the radial baseline $\varphi_u$.}
\begin{align}
	\label{eq:Spin}
	\frac{\Lambda}{d}=\frac{sZ_1^s\pm\sqrt{\pa{Z_1^s}^2+Z_0^sZ_2^s}}{2Z_0^s},
\end{align}
where $s=\pm$ according to the choice of modes in Eq.~\eqref{eq:Z}.

In conclusion, we have provided two ways to extract the offset $\Lambda$, and hence infer black hole spin via Eq.~\eqref{eq:Offset}, from sparse interferometric data.  The first method, using Eq.~\eqref{eq:SpinSolution} with $m=0$ (for instance), allows one to estimate $\Lambda$ after measuring only the six coefficients $P_0^{\mathcal{E},\mathcal{O}}$, $P_1^{\mathcal{E},\mathcal{O}}$, and $P_2^{\mathcal{E},\mathcal{O}}$.  The signature of $\Lambda$ in low-order image modes is promising for measurement of spin with a sparse array.  The second method, using Eq.~\eqref{eq:Spin}, allows for spin to be inferred from visibility measurements along a single radial baseline without requiring absolute phase information (see App.~\ref{app:Visibility}).

\acknowledgements{We acknowledge Jason Dexter, Charles Gammie, Samuel Gralla, and Shahar Hadar for useful conversations.  We thank the National Science Foundation (PHY-1205550) and the Gordon and Betty Moore Foundation (GBMF-5278).  AL also acknowledges the Jacob Goldfield Foundation.  This work was partly completed at the Black Hole Initiative at Harvard University, which is supported by a grant from the John Templeton Foundation.\hfill\scalebox{.01}{\cite{Smolin2017}}}

\appendix

\section{Parallel transport of polarization}
\label{app:Polarization}

The four-momentum $p_\mu$ of a photon in Kerr is given by Eq.~\eqref{eq:KerrMomentum}.  Raising the index and expanding near asymptotic infinity yields the leading-order expansion
\begin{align}
	p^\mu\pd_\mu\approx\pd_t\pm_r\pd_r\pm_\theta\frac{\sqrt{\Theta(\theta)}}{r^2}\pd_\theta+\frac{\lambda}{r^2\sin^2{\theta}}\pd_\phi,
\end{align}
with subleading terms suppressed in inverse powers of $r$.  Thus, a photon received by a distant observer at large radius $r\to\infty$ and inclination $\theta$ has four-momentum
\begin{align}
	p^\mu\pd_\mu=\pd_t+\pd_r+\frac{\beta}{r^2}\pd_\theta-\frac{\alpha}{r^2\sin{\theta}}\pd_\phi,
\end{align}
where we used the definition \eqref{eq:BardeenCoordinates} of the Bardeen coordinates $(\alpha,\beta)$.  In particular,
\begin{align}
	\label{eq:ImpactParameters}
	\alpha=-r^2\sin{\theta}\,p^\phi,\quad
	\beta=r^2\,p^\theta.
\end{align}
Note that $\sign\pa{\alpha}=-\sign\pa{p^\phi}$ and $\sign\pa{\beta}=\sign\pa{p^\theta}$.

At large radius, $p\cdot f\approx-f^t+f^r+\beta f^\theta-\alpha\sin{\theta}f^\phi$, and hence the orthogonality condition \eqref{eq:ParallelTransport} implies that
\begin{align}
	\label{eq:OrthogonalityRelation}
	f^r=f^t-\beta f^\theta+\alpha\sin{\theta}f^\phi.
\end{align}
Using this, the large-radius expansion of Eq.~\eqref{eq:PenroseWalker} yields
\begin{align}
	\kappa_1&\approx\pa{-rf^\theta}\beta+\pa{-r\sin{\theta}f^\phi}\nu,\\
	\kappa_2&\approx\pa{-rf^\theta}\nu-\pa{-r\sin{\theta}f^\phi}\beta.
\end{align}
Inverting these relations yields
\begin{align}
	\label{eq:VerticalPolarizationComponent}
	-rf^\theta&=\frac{\beta\kappa_1+\nu\kappa_2}{\beta^2+\nu^2},\\
	\label{eq:HorizontalPolarizationComponent}
	-r\sin{\theta}f^\phi&=\frac{\nu\kappa_1-\beta\kappa_2}{\beta^2+\nu^2}.
\end{align}

At large radius, $f\cdot f\approx r^2\br{\pa{f^\theta}^2+\sin^2{\theta}\pa{f^\phi}^2}$, and hence a basis for unit-normalized vectors $f\cdot f=1$ is
\begin{align}
	\hat{\beta}=-\frac{1}{r}\pd_\theta&:\quad
	\pa{f^\theta,f^\phi}=\pa{-\frac{1}{r},0},\\
	\hat{\alpha}=\frac{1}{r\sin{\theta}}\pd_\phi&:\quad
	\pa{f^\theta,f^\phi}=\pa{0,\frac{1}{r\sin{\theta}}}.
\end{align}
$\hat{\beta}$ and $\hat{\alpha}$ are locally aligned with $-\pd_\theta$ and $\pd_\phi$, respectively.

On the celestial sphere, the polarization vector is thus
\begin{align}
	f^\theta\pd_\theta+f^\phi\pd_\phi=\pa{-rf^\theta}\hat{\beta}+\pa{r\sin{\theta}f^\phi}\hat{\alpha},
\end{align}
which has vector components on the observer screen
\begin{align}
	\pa{f_\alpha,f_\beta}=\frac{1}{\beta^2+\nu^2}\pa{\beta\kappa_2-\nu\kappa_1,\beta\kappa_1+\nu\kappa_2}.
\end{align}
Since this vector $\vec{f}=f_\alpha\hat{\alpha}+f_\beta\hat{\beta}$ has norm
\begin{align}
	\vec{f}\cdot\vec{f}=\frac{\kappa_1^2+\kappa_2^2}{\beta^2+\nu^2},
\end{align}
the (unit-normalized) polarization direction in the sky is
\begin{align}
	\label{eq:PolarizationDirection}
	\pa{\mathcal{E}_\alpha,\mathcal{E}_\beta}=\frac{\pa{\beta\kappa_2-\nu\kappa_1,\beta\kappa_1+\nu\kappa_2}}{\sqrt{\pa{\kappa_1^2+\kappa_2^2}\pa{\beta^2+\nu^2}}}.
\end{align}

Next, we want to define the angle of the plane of polarization $\chi$ to be increasing in the counterclockwise direction and vanishing when the polarization is vertical (that is, we want $\chi=0$ when $\mathcal{E}_\alpha=0$).  As such, we set
\begin{align}
	\pa{\mathcal{E}_\alpha,\mathcal{E}_\beta}=\pa{-\sin{\chi},\cos{\chi}},
\end{align}
or equivalently,
\begin{align}
	\label{eq:PolarizationAngle}
	\chi=\arctan\pa{-\frac{\mathcal{E}_\alpha}{\mathcal{E}_\beta}}.
\end{align}

Lastly, we define the complex polarization
\begin{align}
	P=Q+iU
	=mIe^{2i\chi},
\end{align}
such that $\updownarrow$ polarization ($\chi=0$) has $Q>0$, $\leftrightarrow$ polarization ($\chi=\pi/2$) has $Q<0$, $\mathrlap{\searrow}{\nwarrow}$ polarization ($\chi=\pi/4$) has $U>0$, and $\mathrlap{\nearrow}{\swarrow}$ polarization ($\chi=3\pi/4$) has $U<0$. Then,
\begin{align}
	\frac{Q}{mI}&=\cos{2\chi}
	=\frac{1-\tan^2{\chi}}{1+\tan^2{\chi}}
	=\mathcal{E}_\beta^2-\mathcal{E}_\alpha^2,\\
	\frac{U}{mI}&=\sin{2\chi}
	=\frac{2\tan{\chi}}{1+\tan^2{\chi}}
	=-2\mathcal{E}_\alpha\mathcal{E}_\beta.
\end{align}
As such,
\begin{align}
	\label{eq:QoverI}
	\frac{Q}{mI}&=\frac{\pa{\beta\kappa_1+\nu\kappa_2}^2-\pa{\nu\kappa_1-\beta\kappa_2}^2}{\pa{\beta^2+\nu^2}\pa{\kappa_1^2+\kappa_2^2}},\\
	\label{eq:UoverI}
	\frac{U}{mI}&=\frac{2\pa{\beta\kappa_1+\nu\kappa_2}\pa{\nu\kappa_1-\beta\kappa_2}}{\pa{\beta^2+\nu^2}\pa{\kappa_1^2+\kappa_2^2}}.
\end{align}
Finally, we obtain
\begin{align}
	\label{eq:PolarizationPW}
	\frac{P}{mI}=\frac{\pa{\beta+i\nu}\pa{\kappa_1-i\kappa_2}}{\pa{\beta-i\nu}\pa{\kappa_1+i\kappa_2}}.
\end{align}
In particular, it follows that the quantity
\begin{align}
	\frac{\bar{\kappa}}{\kappa}=\pa{\frac{\beta-i\nu}{\beta+i\nu}}\frac{P}{mI}
\end{align}
is complex conjugated across subrings in images of reflection-symmetric matter distributions [Eq.~\eqref{eq:SubringsKappa}].

\subsection{Gauge-fixing}

The polarization $f^\mu$ is the Fourier transform of the gauge potential $A^\mu$.  The orthogonality condition \eqref{eq:OrthogonalityRelation} is the momentum-space realization of the Lorenz gauge condition $\nabla_\mu A^\mu=0$.  This condition does not completely fix the gauge, since it still allows for a residual gauge freedom under gauge transformations $A^\mu\to A^\mu+\nabla^\mu\Phi$ such that $\nabla^2\Phi=0$.  To complete the gauge-fixing, one may additionally demand that the harmonic scalar $\Phi(x^\mu)$ be such that $A^t=0$.  The analogous momentum-space statement is that the residual gauge freedom under shifts $f^\mu\to f^\mu+cp^\mu$ (which maintain the Lorenz gauge condition $p\cdot f=0$ while leaving the normalization $f\cdot f$ invariant) may be used to set $f^t=0$.

In this appendix, we never needed to completely fix the gauge by setting $f^t=0$, since the dependence on $f^t$ only enters at subleading order in $1/r$.

\subsection{Consistency checks}

From Eq.~\eqref{eq:PenroseWalker}, one can solve for $\kappa_1$ and $\kappa_2$ to find
\begin{align}
	\label{eq:Kappa1}
	\kappa_1&=-\delta_0f^t+\delta_1f^r+\delta_2f^\theta+\delta_3f^\phi,\\
	\label{eq:Kappa2}
	\kappa_2&=-\gamma_0f^t+\gamma_1f^r+\gamma_2f^\theta+\gamma_3f^\phi,
\end{align}
where
\begin{subequations}
\label{eq:KappaInversion}
\begin{align}
	\delta_0&=r\,p^r+a^2\cos{\theta}\sin{\theta}\,p^\theta,\\
	\delta_1&=r\,p^t-ar\sin^2{\theta}\,p^\phi,\\
	\delta_2&=a^2\sin{\theta}\cos{\theta}\,p^t-a\pa{r^2+a^2}\cos{\theta}\sin{\theta}\,p^\phi,\\
	\delta_3&=ar\sin^2{\theta}\,p^r+a\pa{r^2+a^2}\cos{\theta}\sin{\theta}\,p^\theta,\\
	\gamma_0&=-a\cos{\theta}\,p^r+ar\sin{\theta}\,p^\theta,\\
	\gamma_1&=a^2\cos{\theta}\sin^2{\theta}\,p^\phi-a\cos{\theta}\,p^t,\\
	\gamma_2&=ar\sin{\theta}\,p^t-r\pa{r^2+a^2}\sin{\theta}\,p^\phi,\\
	\gamma_3&=r\pa{r^2+a^2}\sin{\theta}\,p^\theta-a^2\cos{\theta}\sin^2{\theta}\,p^r.
\end{align}
\end{subequations}
When the additional gauge condition $f^t=0$ is imposed, Eqs.~\eqref{eq:KappaInversion} correctly reduce to Eqs.~(14)--(22) of Ref.~\cite{Dexter2016}.


Finally, comparing $\kappa_1$ and $\kappa_2$ for a photon loaded onto the $n^\text{th}$ ray at $\theta_\mathrm{s}$ with their values for a photon loaded onto the $(n+1)^\text{th}$ ray at the reflected angle $\pi-\theta_\mathrm{s}$, one finds using Eqs.~\eqref{eq:MomentumFlip} and \eqref{eq:PolarizationFlip} that $\gamma_0$, $\gamma_1$, $\delta_2$, and $\delta_3$ change sign, while the other quantities $\delta_0$, $\delta_1$, $\gamma_2$, and $\gamma_3$ do not.  Therefore, it follows from Eqs.~\eqref{eq:Kappa1} and \eqref{eq:Kappa2} that $\kappa_2$ also changes sign while $\kappa_1$ remains unchanged, in accordance with Eq.~\eqref{eq:SubringsKappa} in the main text.

\subsection{Comparison with previous works}

Since
\begin{align}
	\label{eq:SphereDirections}
	\hat{\alpha}=\hat{\phi},\quad
	\hat{\beta}=-\hat{\theta},
\end{align}
Eq.~\eqref{eq:PolarizationDirection} agrees with Eq.~(B64) of Ref.~\cite{Li2009}, wherein $\kappa\propto\kappa_1+i\kappa_2$, $(S,T)=(\nu,\beta)$, and $(f^{\hat{\vartheta}},f^{\hat{\varphi}})=(\hat{\theta},\hat{\phi})$.  Our expressions likewise agree with Eq.~(7) of Ref.~\cite{Connors1980}, noting that $\kappa=\kappa_2-i\kappa_1$, $(S,T)=(\nu,\beta)$, and $(X_\infty,Y_\infty)=(\hat{\theta},\hat{\phi})$ therein; Eqs.~\eqref{eq:PolarizationDirection} and \eqref{eq:SphereDirections} are also consistent with the statements made in Ref.~\cite{Dexter2016} below Eq.~(35), noting that $\kappa=\kappa_1-i\kappa_2$, $\gamma=\nu$, and $(\hat{\theta}_0,\hat{\phi}_0)=(\hat{\theta},\hat{\phi})$ therein.

On the other hand, the seminal works by Connors and Stark \cite{Connors1977} and by Connors, Piran, and Stark \cite{Connors1980} contain errors.  The first paper incorrectly claims that both $f^t$ and $f^r$ can be set to zero at infinity, in contradiction with Eq.~\eqref{eq:OrthogonalityRelation} (and there is also a missing factor of $i$ in front of $\kappa_1$ in the definition of $\kappa$).  The second paper appears to claim via Eqs.~(1) and (7) that the Stokes parameters $(Q,U)$ are related to quantities $(X,Y)$ corresponding to our vector components $(\mathcal{E}_a,\mathcal{E}_b)$ in Eq.~\eqref{eq:PolarizationDirection}.  However, this cannot be correct because it would imply that these Stokes parameters change sign when $f^\mu\to-f^\mu$ (and therefore $\kappa_{1,2}\to-\kappa_{1,2}$), unlike our expressions \eqref{eq:QoverI} for $Q$ and \eqref{eq:UoverI} for $U$, which remain invariant under this sign flip (the expected behavior, as they are not vectors).

Finally, Chandrasekhar's book \cite{Chandrasekhar1983} also contains a mistake in \S63(e) (pp358-361).  Noting that $\kappa=\kappa_2+i\kappa_1$, $\gamma=\nu$, and $(\mathcal{E}^\varphi,\mathcal{E}^\theta)=(-\hat{\phi},\hat{\theta})$ therein, his final expressions in Eq.~(265) have an incorrect relative sign, which could for instance be fixed by sending $\gamma\to-\gamma$.

\onecolumngrid

\section{Interferometric calculations}
\label{app:Visibility}

Following the notation of Eq.~\eqref{eq:ThinRing}, let $\mathcal{P}_m$ denote the Fourier coefficients of $\mathcal{P}(\rho,\varphi_u)$.  By Eqs.~\eqref{eq:RelationToP} and \eqref{eq:BetaUpsilon},
\begin{align}
	\label{eq:CoefficientRelations}
	P_m^\mathcal{O}=\frac{d^2}{4}\mathcal{P}_{m-2}-id\Lambda\mathcal{P}_{m-1}-\Lambda^2\mathcal{P}_m,\qquad
	\overline{P_{-m}^\mathcal{E}}=\frac{d^2}{4}\mathcal{P}_{m+2}+id\Lambda\mathcal{P}_{m+1}-\Lambda^2\mathcal{P}_m.
\end{align}
Defining $\tilde{\mathcal{P}}_m=\mathcal{P}_me^{im\varphi_u+im^2\frac{\pi}{2}}$, the amplitudes $P_\pm^\mathcal{O}(\varphi_u)$ corresponding to the odd subrings are
\begin{align}
	P_+^\mathcal{O}(\varphi_u)&\equiv\frac{1}{\pi}\sum_{m=-\infty}^\infty P_m^\mathcal{O}e^{im\varphi_u+im^2\frac{\pi}{2}}
	=\frac{1}{\pi}\sum_{m=-\infty}^\infty\br{\frac{d^2}{4}e^{2i\varphi_u}+(-1)^md\Lambda e^{i\varphi_u}-\Lambda^2}\tilde{\mathcal{P}}_m,\\
	P_-^\mathcal{O}(\varphi_u)&\equiv\frac{1}{\pi}\sum_{m=-\infty}^\infty \pa{-1}^mP_m^\mathcal{O}e^{im\varphi_u+im^2\frac{\pi}{2}}
	=\frac{1}{\pi}\sum_{m=-\infty}^\infty\br{(-1)^m\frac{d^2}{4}e^{2i\varphi_u}-d\Lambda e^{i\varphi_u}-(-1)^m\Lambda^2}\tilde{\mathcal{P}}_m,
\end{align}
whereas for even subrings, the amplitudes $P_\pm^\mathcal{E}(\varphi_u)$ are
\begin{align}
	\overline{P_+^\mathcal{E}(\varphi_u)}&\equiv\frac{1}{\pi}\sum_{m=-\infty}^\infty (-1)^m\overline{P_{-m}^\mathcal{E}}e^{im\varphi_u+im^2\frac{\pi}{2}}
	=\frac{1}{\pi}\sum_{m=-\infty}^\infty\br{(-1)^m\frac{d^2}{4}e^{-2i\varphi_u}+d\Lambda e^{-i\varphi_u}-(-1)^m\Lambda^2}\tilde{\mathcal{P}}_m,\\
	\overline{P_-^\mathcal{E}(\varphi_u)}&\equiv\frac{1}{\pi}\sum_{m=-\infty}^\infty \overline{P_{-m}^\mathcal{E}}e^{im\varphi_u+im^2\frac{\pi}{2}}
	=\frac{1}{\pi}\sum_{m=-\infty}^\infty\br{\frac{d^2}{4}e^{-2i\varphi_u}-(-1)^md\Lambda e^{-i\varphi_u}-\Lambda^2}\tilde{\mathcal{P}}_m.
\end{align}
If we define
\begin{align}
	S_m^\pm&\equiv\frac{1\pm(-1)^m}{\pi},\\
	X^\pm&\equiv P_+^\mathcal{O}(\varphi_u)\pm P_-^\mathcal{O}(\varphi_u)
	=\sum_{m=-\infty}^\infty\br{\pa{\frac{d^2}{4}e^{2i\varphi_u}-\Lambda^2}S_m^\pm\mp d\Lambda e^{i\varphi_u}S_m^\mp}\tilde{\mathcal{P}}_m,\\
	Y^\pm&\equiv\overline{P_+^\mathcal{E}(\varphi_u)}\pm\overline{P_-^\mathcal{E}(\varphi_u)}
	=\sum_{m=-\infty}^\infty\br{\pm\pa{\frac{d^2}{4}e^{-2i\varphi_u}-\Lambda^2}S_m^\pm+d\Lambda e^{-i\varphi_u}S_m^\mp}\tilde{\mathcal{P}}_m,
\end{align}
then by explicit computation, one can check that
\begin{align} 
	\label{eq:QuadraticLambda}
	\frac{1}{4}\pa{X^\pm e^{-2i\varphi_u}\mp Y^\pm e^{2i\varphi_u}}\pm\frac{\Lambda}{d}\pa{X^\mp e^{-i\varphi_u}\mp Y^\mp e^{i\varphi_u}}-\frac{\Lambda^2}{d^2}\pa{X^\pm\mp Y^\pm}=0. 
\end{align}
In terms of the quantities $Z_k^\pm$ defined in Eq.~\eqref{eq:Z}, this relation may be rewritten as
\begin{align}
	Z_0^\pm\frac{\Lambda^2}{d^2}\mp Z_1^\pm\frac{\Lambda}{d}-\frac{Z_2^\pm}{4}=0.
\end{align}
This is a quadratic equation in $\Lambda/d$, whose solution is given in Eq.~\eqref{eq:Spin}.

Because of short coherence times from atmospheric turbulence and stochastic errors in reference oscillators, absolute visibility phase information is generally not accessible with millimeter VLBI \cite{Thompson2017}.  However, after calibration of slowly varying corrections, visibility phases are generally coherent across the recorded bandwidth.  Thus, a single-baseline interferometer with sufficiently wide bandwidth could measure $I_\pm^{\mathcal{E},\mathcal{O}}(\varphi_u)$ up to a single unknown and time-dependent phase $\phi_{\mathcal{E},\mathcal{O}}(t)$.  For instance, with primed/unprimed variables respectively denoting measured/exact quantities,
\begin{align}
	I_\pm^{'\mathcal{E},\mathcal{O}}(\varphi_u)=I_\pm^{\mathcal{E},\mathcal{O}}(\varphi_u)e^{i\phi_{\mathcal{E},\mathcal{O}}(t)}. 
\end{align}

After calibration, phase errors are also generally coherent among different measured Stokes parameters.  Hence, the corresponding coefficients for the polarization images will share the \textit{same} unknown phase:
\begin{align}
	P_\pm^{'\mathcal{E},\mathcal{O}}(\varphi_u)=P_\pm^{\mathcal{E},\mathcal{O}}(\varphi_u)e^{i\phi_{\mathcal{E},\mathcal{O}}(t)}.
\end{align}
As a result, certain combinations of observables are independent of the phase errors.  For instance, 
\begin{align}    
	\frac{P_\pm^{'\mathcal{E},\mathcal{O}}(\varphi_u)}{I_\pm^{'\mathcal{E},\mathcal{O}}(\varphi_u)}=\frac{P_\pm^{\mathcal{E},\mathcal{O}}(\varphi_u)}{I_\pm^{\mathcal{E},\mathcal{O}}(\varphi_u)}.
\end{align}

Because intensity images are self-similar for adjacent subrings [Eq.~\eqref{eq:OnAxisSubringIntensity}], $I_\pm^\mathcal{O}(\varphi_u)=I_\pm^\mathcal{E}(\varphi_u)$.  As such, simultaneous measurements on a pair of collinear baselines sampling an odd and an even subring with associated phase uncertainties $\phi_\mathcal{O}(t)$ and $\phi_\mathcal{E}(t)$ could determine the \textit{phase difference}
\begin{align}
	e^{i\Delta(t)}\equiv e^{i\br{\phi_\mathcal{O}(t)-\phi_\mathcal{E}(t)}}
	=\frac{I_\pm^{'\mathcal{O}}(\varphi_u)}{I_\pm^{'\mathcal{E}}(\varphi_u)}.  
\end{align}
In terms of this measurable phase difference,  Eq.~\eqref{eq:QuadraticLambda} becomes 
\begin{align} 
	\frac{1}{4}\pa{\ab{X^\pm}e^{i\Delta}e^{-2i\varphi_u}\mp\ab{Y^\pm}e^{2i\varphi_u}}\pm\frac{\Lambda}{d}\pa{\ab{X^\mp}e^{i\Delta}e^{-i\varphi_u}\mp\ab{Y^\mp}e^{i\varphi_u}}-\frac{\Lambda^2}{d^2}\pa{\ab{X^\pm}e^{i\Delta}\mp\ab{Y^\pm}}=0.
\end{align}
Thus, the black hole spin could be estimated from $\Lambda$ using a pair of single-baseline interferometers (e.g., a single radial baseline sampled with two widely spaced frequency bands), even in the absence of absolute visibility phase information.  

\bibliographystyle{utphys}
\bibliography{PolarizationRing.bib}

\end{document}